# The role of the coherence length for the establishment of global phase coherence in three-dimensional arrays of ultra-thin quasi-one-dimensional superconducting Pb and NbN nanowires

Chi Ho Wong[1,2*], Frank L. Y. Lam[3*], Junying Shen[1], Minquan He[1†], Xijun Hu[3], and Rolf Lortz[1‡]

[1] *Department of Physics, Hong Kong University of Science and Technology, Clear Water Bay, Kowloon, Hong Kong*

[2] *Institute of Physics and Technology, Ural Federal University, Russia*

[3] *Department of Chemical and Biomolecular Engineering, Hong Kong University of Science and Technology, Clear Water Bay, Kowloon, Hong Kong*

We have fabricated 5 nm ultra-thin NbN nanowires that form a dense and regular array in the linear channels of mesoporous SBA-15 silica substrates. Bulk NbN is a well-known classical superconductor with $T_c$ of 16 K. We show that, by being incorporated into this nanostructure, the composite material exhibits typical quasi-one-dimensional characteristics. We compare the superconducting properties with those of superconducting Pb nanowires of same dimensionality in identical configuration within the linear SBA-15 pores. While Pb nanowire arrays show a pronounced crossover from 1D superconductivity at high temperatures to a 3D bulk superconducting state in the low temperature regime with true zero resistance triggered by transversal Josephson interaction, this transition appears to be completely absent in the NbN nanowire array. The small coherence length in NbN, which strongly suppresses the Josephson coupling is discussed as the origin of this difference.

A quasi one-dimensional (quasi-1D) superconductor is a superconducting material which is confined in a 1D geometry of lateral dimensionality equal to or less than the superconducting coherence length of the corresponding bulk material [1,2]. Such nanowires or linear atomic chains can either be realized in the form of individual free-standing nanowires [2,3], or can be arranged to form an array of parallel 1D nano-elements [1,4-7]. The latter was realized in the form of artificially grown nanostructures [1,5,6,8,9] or found in intrinsic quasi-1D superconducting materials such as $Tl_2Mo_6Se_6$ [10] or $Sc_3CoC_4$ [11,12]. The latter contain many parallel conducting 1D atomic chains separated by either insulating layers or a non-superconducting metallic host. Dependent on the superconducting coherence

---

length and the separation of the chains, they can have a weak transverse coupling via the Josephson or proximity effect, respectively [8,10,13]. The confinement of a superconducting material in a small dimensionality can have some interesting effects [14-22]. The Mermin-Wagner theorem [23,24] predicts that a superconductor in a dimensionality smaller than 3D should never reach a true zero-resistance state at any temperature above the absolute zero point. 2D superconductors have been found that they can overcome this limitation by the famous Berezinski-Kosterlitz-Thouless (BKT) transition [25-27]. However, it is expected that a truly 1D superconductor will only show a kink at the temperature at which the Cooper pairs form, while the resistance drops very continuously to zero without reaching it at finite temperatures [2,7,28-34]. Quasi-1D superconductors are located between the 3D and 1D limits. In the form of 1D nanowires with a thickness of several tens of nanometers, zero resistance can be observed under certain conditions. This is an effect of the finite thickness of the material. Furthermore, in the case of parallel arrays of nanowires or nano-chains, the transverse Josephson coupling can trigger a macroscopic ordering of the phases of the individual superconducting order parameters within the chains, which can mediate a bulk superconducting state over the entire array at low temperatures [8-10,13,35-42]. Such a phase ordering transition has been shown to fall in the same universality class as the BKT transition in a 2D superconducting film [43]. We recently presented a study of an artificially grown array structure of parallel ultra-thin Pb nanowires arranged in close proximity in the pores of an insulating mesoporous SBA-15 substrate (Pb-SBA-15) [8]. We were able to show that this arrangement preserves a zero resistance state, while the confinement in the nanostructure increases the onset critical temperature by almost 4 K (~50% of the bulk $T_c$). In addition, the critical field increased by a factor of ~200 from 800 Oe to more than 14 T. This observation suggested that nano-structuring can be a promising way to improve superconductors to make them more suitable for applications such as high magnetic field solenoids. In this article, we will investigate the role of the superconducting coherence length in such array structures by replacing the Pb nanowires with a superconductor with higher transition temperature and shorter coherence length. $Nb_3Sn$ or NbTi would be interesting options because of their important role in applications. However their growth in a confined nanostructure would be rather difficult. NbN is almost as important in superconducting applications as NbTi and has similar characteristics with a maximum transition temperature of 16 K [44], a short coherence length of 6.5 nm [45] and an upper critical field of 15 T [44]. NbN-SBA-15 can be prepared in a manner similar to Pb-SBA-15, but within an N atmosphere to provide the N source.

*Methods*

NbN-SBA-15 nanowires were prepared in the mesoporous SBA-15 substrates similarly to Pb-SBA-15 [8], using a chemical vapor deposition (CVD) technique, but with an organometallic precursor containing Nb: Tetrakis (2,2,6,6-tetramethyl-3,5-heptanedionato)niobium(IV), 99% [Nb(TMHD)$_4$]. Ammonia was used as a reducing agent, wherein the niobium precursor was converted into an elementary form of niobium in the pores of SBA-15. The Nb then reacts with the nitrogen from the ammonia gas to form NbN. Fig. 1 shows the TEM image of the quasi-1D niobium nitride nanowire arrays produced by the CVD method. The filling factor of the SBA-15 pores is obviously very high and the nanowires are forming a regular dense array of parallel wires. The diameter of the nanowires is ~5nm, identical to that of the Pb nanowires in Pb-SBA-15, since the same SBA-15 mesoporous substrate was used. This allows us to compare the superconducting properties of both composites in detail by taking in account the different superconducting parameters while maintaining the same dimensionality and configuration.

The composition and metal content of the composite were further tested by X-ray fluorescence (XRF) spectroscopy, which gave a molar ratio of Si : Nb of 49.8 : 50.2. The ratio of ~1 further demonstrates the high filling factor of the SBA-15 pores. X-ray photoelectron spectroscopy (XPS) as a surface-sensitive quantitative spectroscopic measuring the element composition was used to verify the metallic nature of the nanowire surfaces.

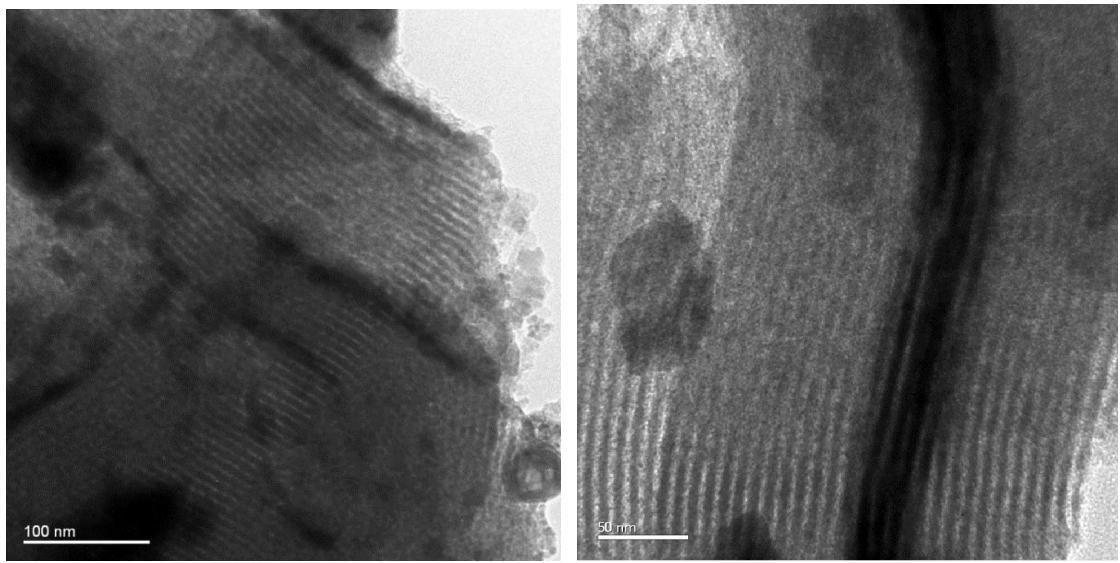

Figure 1. TEM image of quasi-1D NbN nanowire arrays (NbN-SBA15) embedded in the pores of mesoporous SBA-15 substrates (left), compared to a TEM image of Pb nanowire arrays (Pb-SBA15) (right) [8].

The DC magnetization was measured under zero-field (ZFC) and field-cooled (FC) conditions in a Quantum Design Vibrating Sample SQUID magnetometer on a powder sample of 0.5 mg containing many of the small 10 to 100 micrometer SBA-15 crystallites filled with NbN.

The specific heat was measured with a homemade alternating temperature (AC) calorimeter [6,8,46], which provides the required high relative resolution of $\Delta C/C = 10^{-5}$ to resolve the small contribution of the nanowires within the insulating SBA-15 [8]. The measurements were carried out in the temperature range between 2 and 20 K upon cooling with a slow sweep rate of 0.1 K/min, which ensured a high data density. A total mass of 500 µg consisting of a few micrometer-size grains containing the filled SBA-15 crystals and mixed with insulating GE7031 varnish as thermal compound was mounted on the calorimeter chip. The calorimeter was suspended on thin phosphor-bronze wires and contains a resistive Joule heater on the back and a thermocouple connected to the thermal bath. The latter is used to monitor the temperature modulation of the sample using a low-noise preamplifier and a digital lock-in amplifier.

*Magnetization*

Fig. 2a shows zero-field (ZFC) and field-cooled (FC) data of the NbN nanowire arrays in an applied field of 5 Oe. A continuous downturn is observed below 14 K. The ZFC and FC branches strongly deviate from each other and confirm that the down-slope has a superconducting origin. The transition onset at ~14 K is somewhat lower than the bulk $T_c$ value. This is probably due to a nitrogen deficiency in our material that causes a $T_c$ value intermediate between Nb ($T_c$ = 9.25 K [110] and stoichiometric NbN ($T_c$ = 16 K [44]) The transition is obviously much more continuous than in Pb-SBA-15 (shown in Fig. 2b) [8], which is likely a consequence of stronger fluctuations in the phase of the superconducting order parameter due to the shorter coherence length of NbN and thus a stronger 1D character of NbN-SBA-15. Although $T_c$ of NbN is higher, it appears that the superconductivity is weaker and the reason is probably a weaker transverse coupling of the nanowires, thus preserving the 1D nature of these ultra-thin nanowires down to very low temperatures. This contrasts with the dimensional crossover observed for the Pb-SBA-15 nanowires, which originates from a transverse Josephson coupling, which helps establish a 3D bulk phase-coherent superconducting state in macroscopic grains of this composite material [8]. The FC magnetization has not reached a constant value at the lowest accessible temperature, which is another indication that the transverse coupling between the nanowires is very weak. The ZFC magnetization goes through a shallow minimum around 4 K. The origin of it is unknown, but may come from some weak paramagnetic impurity contributions, which are visible because of the small magnitude of the Meissner signal.

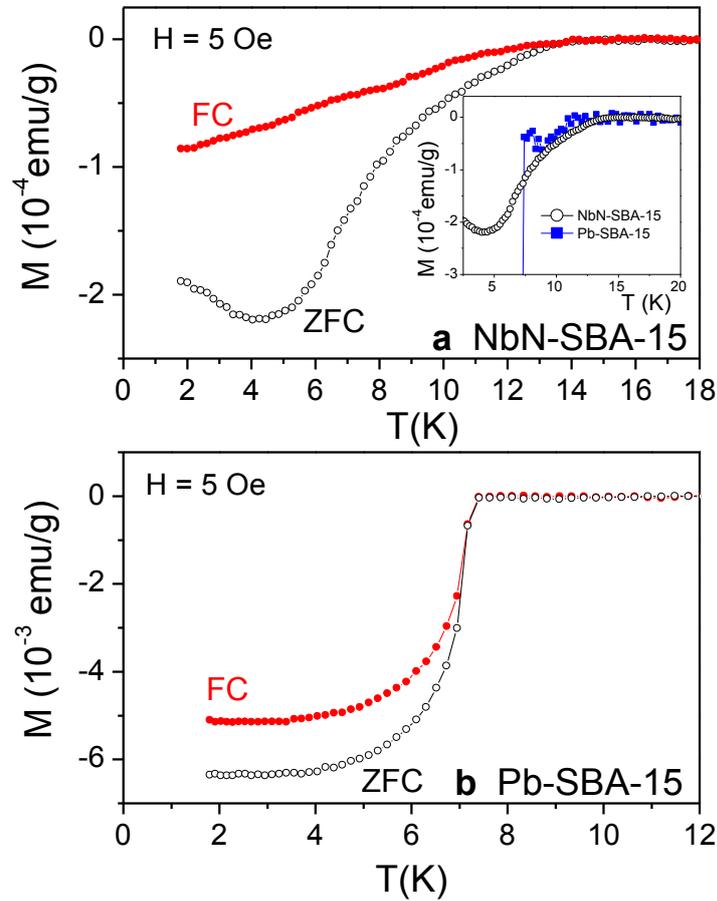

Figure 2. **a)** DC magnetization of NbN-SBA-15 nanowire arrays embedded in mesoporous SBA-15 silica substrates measured in a field of 5 Oe under zero-field (ZFC) and field-cooled (FC) conditions. **b)** Magnetization of Pb-SBA-15 nanowire arrays in mesoporous SBA-15 silica substrates measured under the same conditions as in (**a**) for comparison [8]. The inset of (**a**) shows a zoom on the magnetization data of NbN-SBA-15 and Pb-SBA-15 to show the onset of superconducting fluctuations in the high temperature regime.

In the inset of Fig. 2a we show an enlargement of the transition onsets of NbN-SBA-15 and Pb-SBA-15. Both composites show a start of 1D superconducting fluctuations in the high-temperature regime of similar magnetic magnitude, before the dimensional crossover transition in form of a sharp jump occurs at 7 K in Pb-SBA-15. The latter coincides more or less with the drop of the resistance to zero [8]. This transition is entirely absent in NbN-SBA-15 and the magnitude the magnetization signal remains below $10^{-4}$ emu/g at all temperatures, while the magnetization of Pb-SBA-15 reaches large negative values exceeding -6 x $10^{-3}$ emu/g. This shows that this additional phase-ordering transition, which establishes global phase coherent superconductivity through the entire nanowire array, is absent in NbN-SBA-15, and it remains in the 1D fluctuating superconducting state at all measured finite temperatures.

*Specific heat*

Fig. 3a shows the electronic specific heat as a function of temperature in zero magnetic field and 7 T. This electronic contribution represents only a tiny fraction of ~1% of the total specific heat and was obtained by subtracting data measured in a high field of 14T, which is assumed to suppress the superconductivity largely. The superconducting transition anomaly is quite broad and its shape is similar to that of Pb-SBA-15. It differs significantly from a sharp BCS-type jump-like specific heat transition. In a 7 T magnetic field the transition anomaly is strongly suppressed, although the onset of the transition anomaly is only slightly reduced.

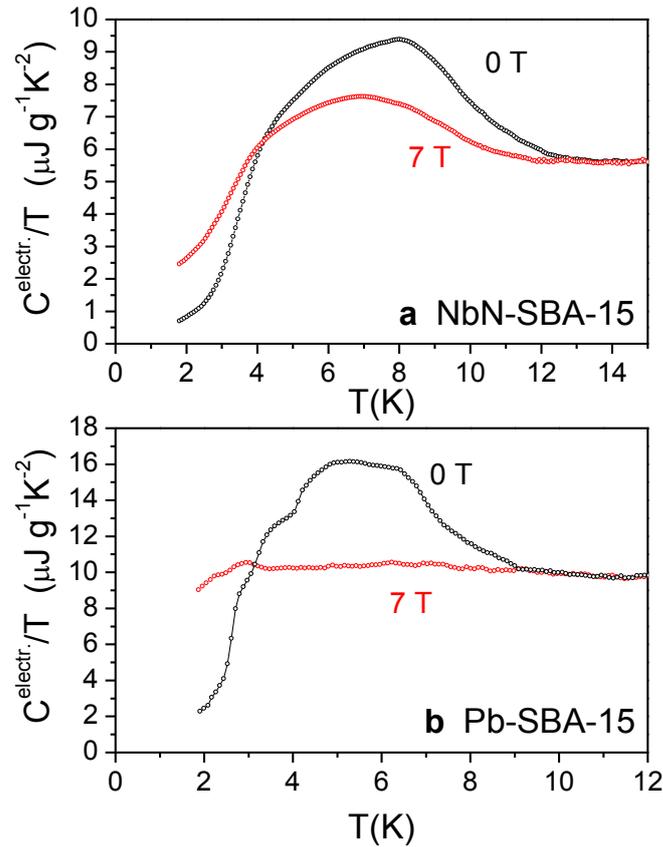

Figure 3. **a)** Approximate electronic contribution of the specific heat of NbN-SBA-15 nanowire arrays in 0 T and 7 T showing a broad superconducting transition anomaly initiated at 13 K. Data measured in 14 T were subtracted to remove the phonon contribution from the NbN and the dominant insulating SBA-15 substrate. **b)** Electronic contribution of the specific heat of Pb-SBA-15 nanowire arrays in 0 T and 7 T measured under identical conditions for comparison [8].

A very similar behavior has been observed for Pb-SBA-15, which is likely a consequence of the 1D nature of superconductivity (see Fig. 3b) [8]. A small fluctuation tail extends up to almost 14 K. The specific heat is a bulk thermodynamic quantity and the large width of the transition shows that superconductivity in the material develops very continuously.

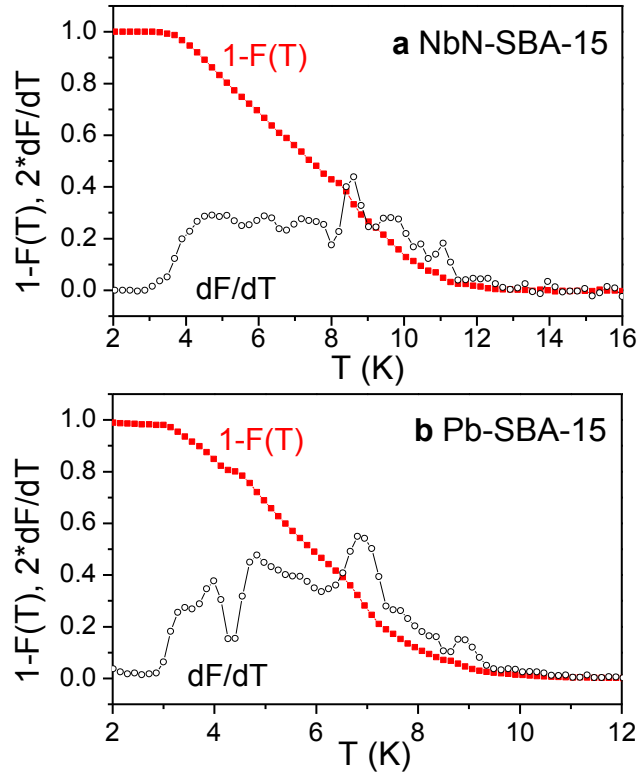

Figure 4. **a)** Superconducting volume fraction of NbN-SBA-15 nanowire arrays $1-F(T)$ and the corresponding $T_c$ distribution $dF/dT$. **b)** $1-F(T)$ and $dF/dT$ of Pb-SBA-15 nanowire arrays for comparison [8].

The zero-field specific heat data allow us to derive the $T_c$ distribution of NbN-SBA-15, as described in detail in Ref. [5,8,47]. In Fig. 4a, $1-F(T)$ (temperature dependence of the superconducting fraction in the sample) and $dF/dT$ ($T_c$ distribution) are shown together with the corresponding data for Pb-SBA-15 (Fig. 4b) [8]. This deconvolution of the specific heat illustrates the large width of the superconducting transition extending from 3.4 K to 13 - 14 K, without any pronounced peaks appearing in $dF(T)/dT$. This confirms the very continuous nature of the superconducting transition in NbN-SBA-15, and together with the magnetization data suggests that the 1D character in this composite material is preserved down into the low-temperature regime without any establishment of global phase coherence in the nanowire arrays.

The additional phase ordering transition which causes the sharper and one order of magnitude larger jump in the magnetization of Pb-SBA-15 is not visible here because such a percolative transition involves only a very small change in the entropy.

*Discussion*

It is noteworthy that the specific heat data of NbN-SBA-15 and Pb-SBA-15 are so similar, although there is a marked difference in the magnetization data with a much sharper transition for Pb-SBA-15. This is due to the fact that the two quantities react sensitively to different effects: the magnetization probes the macroscopic Meissner screening, which is much smaller in NbN-SBA-15 due to the very weak Josephson coupling. The strength gradually increases with the growing coherence length along the nanowires upon lowering of the temperature, while for Pb-SBA-15 the additional phase-ordering transition in the lateral plane initiates a bulk phase-coherent 3D superconducting state with global zero resistance and strong screening currents. The specific heat is mostly sensitive for Cooper pairing and thus demonstrates that the superconducting condensate develops very continuously in a very similar manner in the two materials.

The coherence length of bulk NbN is 6.5 nm [45], which is of the order of the 5 nm diameter of our nanowires, compared to the much longer 83 nm coherence length of bulk Pb. The long coherence length of Pb exceeds the inter-wire distance by an order of magnitude and thus strongly couples the nanowires in the lateral dimension via the Josephson effect. ForNbN-SBA-15 the coherence length is barely of the order of the nanowire separation.

In the following, we will roughly estimate the order of magnitude of the phase ordering transition temperature of the NbN nanowire arrays. Since NbN is a Type II BCS material, the free electron concentration can be estimated from Eq. 1 and 2:

$$\xi_0 = \frac{hv_F}{4\pi E_g} \qquad (1)$$

$$\epsilon_F = \frac{\hbar^2}{2m}\left(\frac{3\pi^3 N}{V}\right)^{\frac{2}{3}} \qquad (2)$$

where $v_F$ is Fermi velocity, $\epsilon_F$ is Fermi level, $E_g$ is the energy gap of superconductor. The carrier concentration is 7.5 x $10^{25}$ m$^{-3}$ and is close to the experimental bulk value of 2 x $10^{26}$ m$^{-3}$ [48]. Related to the short coherence length $\xi_0$ = 6.5nm of NbN, the carrier concentration of the NbN nanowires is three orders of magnitude smaller than that of the Pb nanowires in Pb-SBA-15, which implies that the Josephson coupling of the NbN nanowire arrays is also three orders of magnitude weaker. In addition, the dimensional transition temperature is proportional to the coupling force. We therefore estimate that the dimensional transition temperature of NbN arrays should be in range of milli-Kelvin compared to the ~7 K of Pb-SBA-15. This explains the far more continuous magnetization drop in Fig. 2 and the order of magnitude smaller magnetization of Nb-SBA-15 compared to Pb-SBA-15.

*Conclusions*

We have previously found that in 5 nm ultra-thin Pb nanowires arranged in form of a parallel array within mesoporous SBA-15 silica single crystals, a dimensional crossover from a 1D phase-incoherent fluctuating superconducting state to a 3D bulk superconducting state occurs, characterized by global phase coherence throughout the macroscopic nanowire array. The much more continuous magnetic transition, and the one order of magnitude smaller Meissner magnetization of identically arranged 5 nm NbN nanowires in the pores of SBA-15, suggest that the NbN nanowire arrays remain in the 1D fluctuating state to the lowest temperature of our experiments. Although bulk NbN has far more favorable properties for superconducting applications, such as a higher $T_c$ and critical field, it appears to be less promising for applications in the form of ultra-thin nanowire arrays due to its much shorter coherence length $\xi_0 = 6.5$ nm. This strongly suppresses transversal Josephson coupling of adjacent nanowires within the array. While the nanostructured Pb has shown the potential to remain a superconductor with zero resistance and improved superconducting parameters [8], only a 1D fluctuating superconducting state is found in NbN. Our nanostructured NbN is therefore not a useful superconductor in the technical sense, although phase-incoherent Cooper pairing with finite resistance certainly exists up to 14 K. This reveals the limit of the application of such a nano-structuring technique to applied superconductivity for classical superconductors with higher critical temperatures than Pb, since most of the classical type-II 'old high temperature superconductors' such as NbTi, $V_3Si$ or $Nb_3Sn$ have rather short superconducting coherence lengths.

*Acknowledgments*

R. L. thanks N. Ip, Y. Wang and M. Altman for their continuous financial support and encouragement and U. Lampe for technical support. This work was partially financially supported by the Innovative Exploratory Grant of Hong Kong University of Science and Technology (No. IEG14EG02PG).

**References**

[1] K. Xu, K. Heath, Controlled Fabrication and Electrical Properties of Long Quasi-One-Dimensional Superconducting Nanowire Arrays. *Nano Lett.* **8**, 136-141 (2008).

[2] M. Tian, J. Wang, J. S. Kurtz, Y. Liu, M. H. W. Chan, T. S. Mayer, T. E. Mallouk, Dissipation in Quasi-One-Dimensional Superconducting Single-Crystal Sn Nanowires. *Phys. Rev. B* **71**, 104521-104527 (2005).

[3] J. Wang, Y. Sun, M. Tian, B. Liu, M. Singh, M. H. W. Chan, Superconductivity in Single Crystalline Pb Nanowires Contacted by Normal Metal Electrodes. *Phys. Rev. B* **86**, 035439 (2012).

[4] S. Michotte, L. Piraux, S. Dubois, F. Pailloux, G. Stenuit, J. Govaerts, Superconducting Properties of Lead Nanowires Arrays. *Phys. C* **377**, 267-276 (2002).

[5] W. Shi, Z. Wang, Q. Zhang, Y. Zheng, C. Ieong, M. He, R. Lortz, Y. Cai, N. Wang, T. Zhang, H. Zhang, Z. K. Tang, P. Sheng, H. Muramatsu, Y. A. Kim, M. Endo, P. T. Araujo, M. S. Dresselhaus, Superconductivity in Bundles of Double-Wall Carbon Nanotubes. *Sci. Rep.* **2**, 625 (2012).

[6] R. Lortz, Q. Zhang, W. Shi, J. Ye, C. Qiu, Z. Wang, H. He, P. Sheng, T. Qian, Z. K. Tang, N. Wang, X. Zhang, J. Wang, C. T. Chan, Superconducting Characteristics of 4-Å Carbon Nanotube-Zeolite Composite. *PNAS* **106**, 7299-7303 (2009).

[7] M. Tian, J. Wang, J. Snyder, J. Kurtz, Y. Liu, P. Schiffer, T. E. Mallouk, M. H. W. Chan, Synthesis and characterization of superconducting single-crystal Sn nanowires. *App. Phys. Lett.* **83**, 1620 – 1622 (2003).

[8] M. He, C. H. Wong, P. L. Tse, Y. Zheng, H. Zhang, F. L. Y. Lam, P. Sheng, X. Hu, R. Lortz, "Giant" enhancement of the upper critical field and fluctuations above the bulk $T_c$ in superconducting ultrathin lead nanowire arrays. *ACS Nano* **7**, 4187-4193 (2013).

[9] Y. Zhang, C. H. Wong, J. Shen, S. T. Sze, B. Zhang, H. Zhang, Y. Dong, H. Xu, Z. Yan, Y. Li, X. Hu, R. Lortz, Dramatic enhancement of superconductivity in single-crystalline nanowire arrays of Sn Scientific Reports 6, Article number: 32963 (2016).

[10] B. Bergk, A. P. Petrović, Z. Wang, Y. Wang, D. Salloum, P. Gougeon, M. Potel, R. Lortz, Superconducting Transitions of Intrinsic Arrays of Weakly Coupled One-Dimensional Superconducting Chains: The Case of the Extreme Quasi-1D Superconductor $Tl_2Mo_6Se_6$. *New J. Phys*. **13**, 103018 (2011).

[11] E.-W. Scheidt, C. Hauf, F. Reiner, G. Eickerling, W. Scherer, Possible Indicators for Low Dimensional Superconductivity in the Quasi-1D Carbide $Sc_3CoC_4$. *J. Phys. Conf. Ser.* **273**, 012083-012086 (2011).

[12] M. He, D. Shi, P. L. Tse, C. H. Wong, O. Wybranski, E.-W. Scheidt, G. Eickerling, W. Scherer, P. Sheng, R. Lortz, 1D to 3D dimensional crossover in the superconducting transition of the quasi-one-dimensional carbide superconductor $Sc_3CoC_4$, *J. Phys.: Cond. Mat.* **27**, 075702 (2015).

[13] Z. Wang, W. Shi, H. Xie, T. Zhang, N. Wang, Z. K. Tang, X. X. Zhang, R. Lortz, P. Sheng, I. Sheikin, A. Demuer, Superconducting resistive transition in coupled arrays of 4 Å Carbon Nanotubes. *Phys. Rev. B* **81**, 174530-174539 (2010).

[14] J. H. P. Watson, Transition Temperature of Superconducting Indium, Thallium, and Lead Grains. *Phys. Rev. B* **2**, 1282-1286 (1970).

[15] W.-H. Li, C. C. Yang, F. C. Tsao, S. Y. Wu, P. J. Huang, M. K. Chung, Y. D. Yao, Enhancement of Superconductivity by the Small Size Effect in In Nanoparticles. *Phys. Rev. B*, **72**, 214516-214520 (2005).

[16] E. V. Charnaya, C. Tien, K. J. Lin, C. S. Wur, Yu. A. Kumerzov, Superconductivity of Gallium in Various Confined Geometries. *Phys. Rev. B* **58**, 467-472 (1998).

[17] J. Hagel, M. T. Kelemen, G. Fischer, B. Pilawa, J. Wosnitza, E. Dormann, H. v. Löhneysen, A. Schnepf, H. Schnöckel, U. Neisel, J. Beck, Superconductivity of a Crystalline $Ga_{84}$-Cluster Compound. *J. Low Temp. Phys.* **129**, 133-142 (2002).

[18] K. Ohshima, T. Fujita, Enhanced Superconductivity in Layers of Ga Fine Particles. *J. Phys. Soc. Jpn.* **55**, 2798-2802 (1986).

[19] J. Wang, X. Ma, S. Ji, Y. Qi, Y. Fu, A. Jin, L. Lu, C. Gu, X. C. Xie, M. Tian, J. Jia, Q. Xue, Magnetoresistance Oscillations of Ultrathin Pb Bridges. *Nano Res.* **2**, 671-677 (2009).

[20] J. Wang, X. C. Ma., L. Lu., A. Z. Jin, C. Z. Gu, X. C. Xie, J. F. Jia, X. Chen, Q. K. Xue, Anomalous Magnetoresistance Oscillations and Enhanced Superconductivity in Single-Crystal Pb Nanobelts. *Appl. Phys. Lett.* **92**, 233119 (2008).

[21] Y.-J. Hsu, S.-Y. Lu, Lin, Y.-F. Nanostructures of Sn and Their Enhanced, Shape-Dependent Superconducting Properties. *Small* **2**, 268-273 (2006).

[22] J. Wang, Y. Sun, M. Tian, B. Liu, M. Singh, M. H. W. Chan, Superconductivity in Single Crystalline Pb Nanowires Contacted by Normal Metal Electrodes. *Phys. Rev. B* **86**, 035439 (2012).

[23] P. C. Hohenberg, Existence of Long Range Order in One and Two Dimensions. *Phys. Rev.* **158**, 383-386 (1967).

[24] N. D. Mermin, H. Wagner, Absence of Ferromagnetism or Antiferromagnetism in One- or Two-Dimensional Isotropic Heisenberg Models. *Phys. Rev. Lett.* **17**, 1133-1136 (1966).

[25] J. M. Kosterlitz, J. M. and D. J. Thouless, Ordering, metastability and phase transitions in two-dimensional systems. *J. Phys. C* **6,** 1181-1203 (1973).

[26] J. M. Kosterlitz, The critical properties of the two-dimensional xy model. *J. Phys. C* **7**, 1046-1060 (1974).

[27] V. L. Berezinskii, Destruction of long-range order in one-dimensional and two-dimensional systems having a continuous symmetry group I. Classical systems. *Zh. Eksp. Teor. Fiz.* **59**, 207 (1970) [*Sov. Phys. JETP* **32**, 493-500 (1971)].

[28] J. S. Langer, V. Ambegaokar, Intrinsic Resistive Transition in Narrow Superconducting Channels. *Phys. Rev.* **164**, 498-510 (1967).

[29] D. E. McCumber, B. I. Halperin, Time Scale of Intrinsic Resistive Fluctuations in Thin Superconducting Wires. *Phys. Rev. B* **1**, 1054-1070 (1970).

[30] C. N. Lau, N. Markovic, M. Bockrath, A. Bezryadin, M. Tinkham, Quantum phase slips in superconducting nanowires, *Phys. Rev. Lett.* **87**, 217003 (2001).

[31] N. Giordano, Evidence for macroscopic quantum tunneling in one-dimensional superconductors. *Phys. Rev. Lett.* **61**, 2137-2140 (1988).

[32] R. P. Barber, Jr. and R. C. Dynes, Low-field magnetoresistance in granular Pb films near the insulator-superconductor transition. *Phys. Rev. B* **48**, 10618-10621 (1993).

[33] A. V. Herzog, P. Xiong, F. Sharifi, R. C. Dynes, Observation of a discontinuous transition from strong to weak localization in 1D granular metal wires. *Phys. Rev. Lett.* **76**, 668-671 (1996).

[34] A. V. Herzog, P. Xiong, R. C. Dynes, Magnetoresistance oscillations in granular Sn wires near the superconductor-insulator transition. *Phys. Rev. B* **58**, 14199-14202 (1998).

[35] B. Stoeckly, D. J. Scalapino, Statistical Mechanics of Ginzburg-Landau Fields for Weakly Coupled Chains. *Phys. Rev. B* **11**, 205-210 (1975).

[36] D. J. Scalapino, Y. Imry, P. Pincus, Generalized Ginzburg-Landau Theory of Pseudo-One-Dimensional Systems. *Phys. Rev. B* **11**, 2042-2048 (1975).

[37] K. Kobayashi, D. Stroud, Theory of Fluctuations in a Network of Parallel Superconducting Wires. *Phys. C* **471**, 270-276 (2011).

[38] H. J. Schulz, C. Bourbonnais, Quantum Fluctuations in Quasi-One-Dimensional Superconductors. *Phys. Rev. B* **27**, 5856-5859 (1983).

[39] L. P. Gorkov, I. E. Dzyaloshinskii, Possible Phase Transitions in Systems of Interacting Metallic Filaments (Quasiunidimensional Metals). *Zh. Eksp. Teor. Fiz.* **67**, 397-417 (1974).

[40] R. A. Klemm, H. Gutfreund, Order in Metallic Chains. II. Coupled Chains. *Phys. Rev. B* **14**, 1086-1102 (1976).

[41] P. A. Lee, T. M. Rice, R. A. Klemm, Role of interchain coupling in linear conductors. *Phys. Rev. B* **15**, 2984-3002 (1977).

[42] K. B. Efetov, A. I. Larkin, Effect of fluctuations on the transition temperature in quasi-one dimensional superconductors. *Sov. Phys. JETP* **41**, 76-79 (1975).

[43] M. Y. Sun, Z. L. Hou, T. Zhang, Z. Wang, W. Shi, R. Lortz, P. Sheng, Dimensional crossover transition in a system of weakly coupled superconducting nanowires. *New J. Phys.* **14**, 103018 (2012).

[44] R. J. Donelly, Cryogenics. IN Physics Vade Mecum, ed. H. L. Anderson, American Institute of Physics, New York (1981).

[45] F. Altomare, A. M. Chang, One-Dimensional Superconductivity in Nanowires, John Wiley & Sons, (2013)

[46] P. F. Sullivan, G. Seidel, Steady-State, AC-Temperature Calorimetry. *Phys. Rev.* **17**, 679 – 685 (1968).

[47] Y. Wang, C. Senatore, V. Abächerli, D. Uglietti, R. Flükiger, Specific Heat of $Nb_3Sn$ Wires. *Supercond. Sci. Technol.* **19**, 263-266 (2006).

[48] N. D. Kuz'michev, G. P. Motulevich, P. N. Lebedev, Determination of electronic characteristics of niobium nitride by an optical method, Physics Institute, USSR Academy of Sciences (1983).